\documentclass[twocolumn]{article}
\usepackage{times,bm}
\usepackage{amsmath}
\usepackage[affil-it,]{authblk}
\usepackage[pagebackref=false,colorlinks=true,linkcolor=acsblue,citecolor=olive,urlcolor=acsblue]{hyperref}

\usepackage{geometry}
\usepackage[switch*, displaymath,pagewise, mathlines]{lineno}
\usepackage{graphicx}
\usepackage{cite}
\usepackage{amsmath}
\usepackage{amsfonts}
\usepackage{amssymb}
\usepackage{slashed}
\usepackage{subfloat}
\usepackage{slashed}
\usepackage{color}
\usepackage{float}
\usepackage{multirow,multicol}
\usepackage{tabulary}
\usepackage[flushleft]{threeparttable}
\usepackage{pgfplots,subfigure}
\usepackage{xcolor}
\numberwithin{equation}{section}
\usepackage{tikz}
\usepackage{ulem}
\usepackage{hyperref}
\usepackage{nameref}
\usepackage{comment}

\usepgflibrary{arrows}
\usetikzlibrary{shapes.callouts}
\tikzset{
	level/.style   = { thick, },
	connect/.style = { dotted, red   },
	notice/.style  = { draw, rectangle callout, callout relative pointer={#1} },
	label/.style   = { text width=1cm }
}

\definecolor{acsblue}{RGB}{17,76,139}
\definecolor{shadecolor}{RGB}{255,241,204}
\usepackage{letltxmacro}
\makeatletter
\let\oldr@@t\r@@t
\def\r@@t#1#2{%
	\setbox0=\hbox{$\oldr@@t#1{#2\,}$}\dimen0=\ht0
	\advance\dimen0-0.2\ht0
	\setbox2=\hbox{\vrule height\ht0 depth -\dimen0}%
	{\box0\lower0.4pt\box2}}
\LetLtxMacro{\oldsqrt}{\sqrt}
\renewcommand*{\sqrt}[2][\ ]{\oldsqrt[#1]{#2}}
\makeatother
\usepackage{environ}

\begin{document}

\newcommand{{\ri}}{{\rm{i}}}
\newcommand{{\Psibar}}{{\bar{\Psi}}}
\newcommand*\var{\mathit}

\fontsize{8}{9}\selectfont

\title{\mdseries{Ray geodesics and wave propagation on the Beltrami surface: Optics of an optical wormhole}}

\author{ \textit {\mdseries{Semra Gurtas Dogan}}$^{\ 1}$\footnote{\textit{ E-mail: semragurtasdogan@hakkari.edu.tr (Corr. Author)} }~,~ \textit {\mdseries{Abdullah Guvendi}}$^{\ 2}$\footnote{\textit{E-mail: abdullah.guvendi@erzurum.edu.tr } }~,~ \textit {\mdseries{Omar Mustafa}}$^{\ 3}$\footnote{\textit{ E-mail: omar.mustafa@emu.edu.tr} }  \\
	\small \textit {$^{\ 1}$ \footnotesize Department of Medical Imaging Techniques, Hakkari University, 30000, Hakkari, Türkiye}\\
	\small \textit {$^{\ 2}$\footnotesize  Department of Basic Sciences, Erzurum Technical University, 25050, Erzurum, Türkiye}\\
	\small \textit {$^{\ 3}$\footnotesize  Department of Physics, Eastern Mediterranean University, 99628, G. Magusa, north Cyprus, Mersin 10 - Türkiye}}
\date{}
\maketitle

\begin{abstract}
This study investigates ray geodesics and wave propagation on the Beltrami surface, with a particular emphasis on the effective potentials governing photon dynamics. We derive the geodesic equations and analyze the Helmholtz equation within this curved geometry, revealing that the resulting potentials are purely repulsive. For ray trajectories, the potential is determined by wormhole parameters such as the throat radius (\(\ell\)), radial optical distance (\(u\)), scale parameter (\(R\)), and the angular momentum of the test field. Near the wormhole throat, the potential remains constant, preventing inward motion below a critical energy threshold, whereas at larger radial distances, it decays exponentially, allowing free propagation. In the context of wave propagation, the potential exhibits a centrifugal barrier along with a constant repulsive term at large \(u\). The Beltrami surface, characterized by constant negative Gaussian curvature, serves as a model for graphene sheets and optical wormholes in condensed matter systems. These results allow us to determine the space- and frequency-dependent refractive index of the medium, providing a coherent framework for understanding photon behavior in such systems, with promising implications for material applications.
\end{abstract}

\begin{small}
\begin{center}
\end{center}
\end{small}



\section{\mdseries{Introduction}}\label{sec1}

Ray and wave optics in curved spaces study the propagation of light through non-flat spacetimes, where curvature influences both ray and wave behaviors \cite{1,2}. In ray optics, light follows geodesics-curved paths in spacetime-resulting in phenomena such as gravitational lensing. The deflection of light by structures like wormholes has been analyzed using the Gauss-Bonnet theorem, demonstrating how spacetime geometry affects light trajectories \cite{3,4,5,6}. In wave optics, the Helmholtz wave equation is modified for curved spacetime, altering wave behavior accordingly \cite{7,8}. Depending on the curvature, the refractive index of such surfaces can become complex, indicating light attenuation under specific conditions. Quantum electrodynamics in curved spacetime also offers insights into the refractive index and Green's functions \cite{9,10,11,12,13,14,15,16}. Optical wormholes \cite{17,18,19,20,21}-theoretical constructs that replicate certain characteristics of astrophysical wormholes-employ materials with specially engineered refractive index profiles, such as metamaterials, to guide light in a way analogous to "actual" wormholes \cite{12}. Examples include hollow disclinations in liquid crystal films, which create structures that channel light similarly to conical or anti-conical spacetime-generated wormholes \cite{8}. The study of ray and wave optics in curved spaces, particularly through optical wormholes, deepens our understanding of light in non-flat geometries and has practical applications in novel optical devices, gravitational lensing, and theoretical physics \cite{12}.

\vspace{0.15cm}
\setlength{\parindent}{0pt}

Quantum mechanics on curved surfaces studies how quantum behavior is affected by geometry, leading to phenomena such as quantized energy levels, geometric phases, and forces driven by curvature. The curvature of surfaces such as tori and catenoids alters our understanding of quantum particle dynamics \cite{17,18,19,20,21,22,23,23-b}. Two-dimensional materials, especially graphene and phosphorene, are important in condensed matter physics due to their unique electronic properties and geometric flexibility, serving as analogs for high-energy physics, where electronic states respond to geometry \cite{17,18,19,20,21,22,23,23-b}. Toroidal surfaces are studied in areas such as nanoelectronics, biosensors, and quantum computing. Research has provided exact solutions for quantum particles on toroids, including spin-1/2 particles in external fields and Dirac fermions with varying Fermi velocities, supporting potential qubit applications in graphene (see \cite{23}). The catenoid, a minimal surface similar to a condensed matter wormhole, affects charge carrier dynamics through curvature-induced potentials \cite{22}. Studies on condensed matter wormholes show how position-dependent mass influences the behavior of electron-hole pairs \cite{22}. Despite progress, the Beltrami surface, with its constant negative Gaussian curvature and non-Euclidean geometry, remains less studied \cite{23,23-b}. While recent works have addressed event horizons, Hawking radiation, and Dirac fermions on surfaces of negative Gaussian curvature \cite{25,26,27}, no results have been announced for ray geodesics and wave optics on Beltrami surfaces.

\vspace{0.15cm}
\setlength{\parindent}{0pt}

This paper investigates the effect of a Beltrami wormhole, also called an optical wormhole, on arbitrary ray geodesics and wave dynamics. This work is structured as follows: Section \ref{sec:2} introduces the Beltrami surface, Section \ref{sec:3} examines ray geodesics, Section \ref{sec:4} explores the impact of the curved surface on wave dynamics, and Section \ref{sec:5} discusses the findings.

\begin{figure}[ht]
\centering
\includegraphics[scale=0.40]{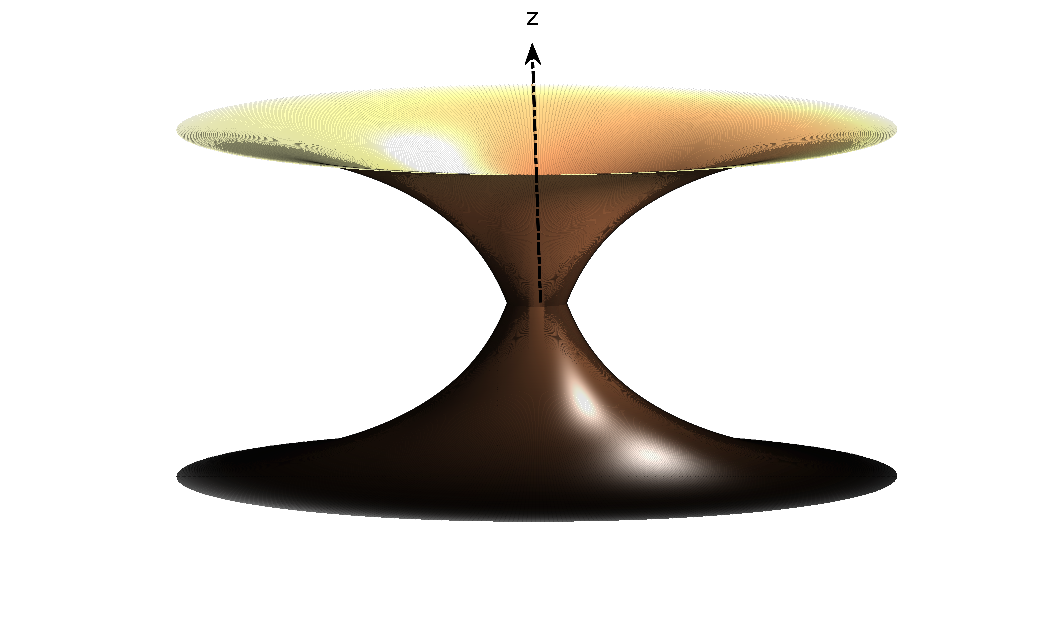}
\caption{\footnotesize 3D plot of a Beltrami wormhole.}
\label{fig:3DWH}
\end{figure}

\section{\mdseries{Beltrami Wormhole} }\label{sec:2}

Among surfaces with negative curvature, those with constant negative Gaussian curvature play a crucial role. When embedded in \(\mathbb{R}^3\), these surfaces exhibit essential singularities \cite{21,25,26}. As a result, it is not possible to represent the entire Lobachevskian geometry on a real two-dimensional surface, forcing us to limit the mapping to a suitable strip of hyperbolic space \cite{21,25,26}. Another important consideration is the explicit parametrization, which allows hyperbolic geometry, along with its singular boundaries, to be expressed in terms of well-defined coordinates. In this context, the line element of any surface with constant negative Gaussian curvature can be reduced to the forms of Beltrami, hyperbolic, or elliptic pseudospheres \cite{21,25,26}. The Beltrami surface offers significant advantages: it can be parametrized using smooth, single-valued functions, and it has a singular boundary corresponding to a maximal circle. Below, we provide an embedding for the Beltrami pseudosphere in three-dimensional space, along with the explicit parametrization in terms of surface coordinates \cite{25}:
\begin{equation}
\begin{split}
&x(u,\phi) = \ell\, e^{\frac{u}{R}} \cos(\phi), \quad y(u,\phi) = \ell \, e^{\frac{u}{R}} \sin(\phi), \\
&z(u) = \pm R \left[ \tan^{-1} f - f\right], \label{P-coordinates}
\end{split}
\end{equation}
where $f(u) = \sqrt{1 - \left(\frac{\ell}{R} e^{\frac{u}{R}} \right)^2}$.Upon inspection, it is evident that the parameterized Beltrami surface exists for \( u \in \left(-\infty, R \log \frac{R}{\ell}\right) \) \cite{25}. The surface can be embedded within \( \mathbb{R}^3 \) and is well-defined over its non-singular portion, with the singular boundary being the maximal circle of radius \( R \), corresponding to the limit value \( u_{\text{max}} = R \log \frac{R}{\ell} \) \cite{25}. Moreover, the equations for the embedding are expressed in terms of analogs of cylindrical coordinates \( u \) and \( \phi \), facilitating navigation along the "meridian" and "parallel" of the surface. Each coordinate is represented by a smooth, well-behaved, single-valued function. These considerations lead to the physical assumption of introducing a limiting value for the surface's (parallel) radius, \( r = \ell\, e^{u/R} \) \cite{25}. The line element describing the Beltrami pseudosphere, as defined in the embedding equation, is given by \cite{21,25}:
\begin{equation}
ds^2 = -c^2 dt^2 + du^2 + \ell^2 e^{\frac{2u}{R}} d\phi^2,\label{metric}
\end{equation}
with the Ricci scalar obtained as: \( \mathcal{R} = -\frac{2}{R^2} \), which is precisely twice the Gaussian curvature \( K = -\frac{1}{R^2} \) \cite{28,29}.

\section{\mdseries{Ray optics} }\label{sec:3}

\begin{figure}[ht]
\centering
\includegraphics[scale=0.50]{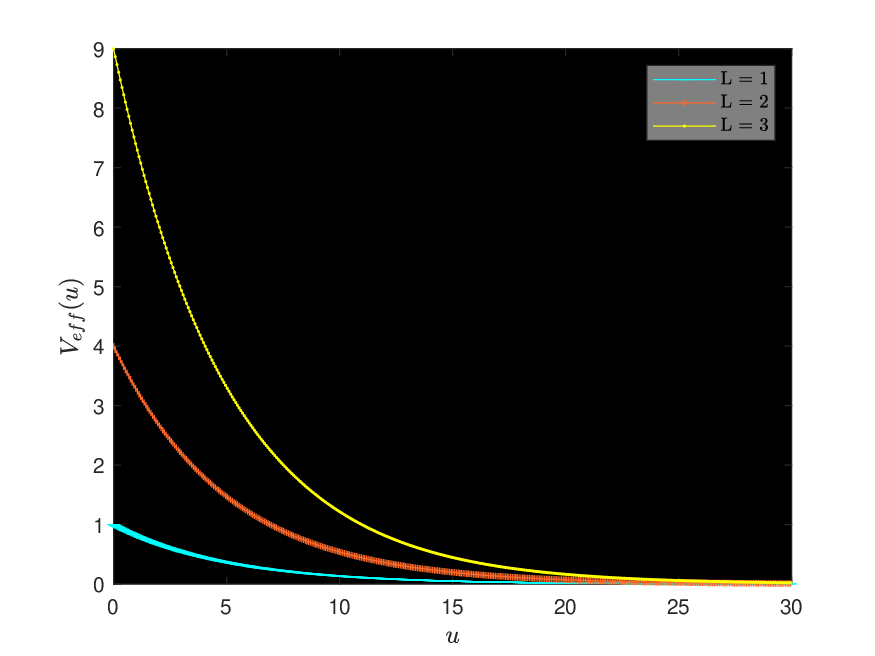}
\caption{\footnotesize A plot of the effective potential \( V_{\text{eff}}(u) \) versus \( u \) is shown for different values of the angular momentum \( L \), with the parameters \( R = 10 \) and \( \ell = 1 \) held constant. The potential exhibits an exponential decay with increasing \( u \) and demonstrates a clear dependence on \( L \). The curves correspond to \( L = 1, 2, 3 \), highlighting how the angular momentum influences both the shape and the magnitude of the potential.}
\label{fig:eff-pot-1}
\end{figure}

In this section, we analyze ray geodesics on the Beltrami surface and determine exact angular trajectories for photons. Let us start by considering the Lagrangian in the following form \cite{1,7,8}:
\begin{equation}
\mathcal{L} = g_{\mu \nu} \frac{dx^{\mu}}{d\lambda} \frac{dx^{\nu}}{d\lambda},
\end{equation}
where \(\lambda\) is the affine parameter of the curve, which serves as the parameter along the path of a particle or light ray. In the context of general relativity, this affine parameter plays a crucial role in parametrizing the motion of particles along geodesics. Geodesics represent the natural trajectories followed by free-falling particles and are the paths that extremize the proper time or the distance in spacetime, depending on the nature of the geodesic. The geodesics of spacetime are determined by solving the Euler-Lagrange equation, which is derived from the above Lagrangian. The general form of this equation is given by \cite{7,8}:
\begin{equation}
\frac{\partial \mathcal{L}}{\partial x^{\mu}} - \frac{d}{d\lambda} \left( \frac{\partial \mathcal{L}}{\partial \dot{x}^{\mu}} \right) = 0.
\end{equation}
This equation dictates how the coordinates \(x^\mu(\lambda)\), where \(\mu\) runs over the space coordinates ($u,\phi$), evolve with respect to the affine parameter \(\lambda\). The Euler-Lagrange equation ensures that the trajectory of a particle or light ray is such that the spacetime interval (or proper time, for timelike geodesics) is extremized. Now, we set \(\mathcal{L} = \kappa\), where \(\kappa\) is a constant that characterizes the type of geodesic we are dealing with. This is achieved by taking the speed of light (\(c\)) in vacuum as unity, \(c = 1\). The constant \(\kappa\) determines whether the geodesic is lightlike, timelike, or spacelike \cite{7,8}. Specifically, we have that \(\kappa = 0\) corresponds to lightlike geodesics, which describe the trajectories of massless particles such as photons, while \(\kappa = -1\) corresponds to timelike geodesics, which describe the paths followed by massive particles \cite{7,8}. Using the line element in Eq.~\eqref{metric}, the Lagrangian takes the form:
\begin{equation}
\mathcal{L} = -\dot{t}^2 + \dot{u}^2 + \ell^2e^{\frac{2u}{R}} \dot{\phi}^2,
\end{equation}
where the dot denotes differentiation with respect to the affine parameter \(\lambda\). Since the Lagrangian \(\mathcal{L}\) does not explicitly depend on \(t\) and \(\phi\), their conjugate momenta are conserved. The Euler-Lagrange equations for \(t\) and \(\phi\) give rise to the following conserved quantities \cite{7,8}:
\begin{equation}
\mathcal{E} = \dot{t}, \quad L = \ell^2e^{\frac{2u}{R}} \dot{\phi}\,,
\end{equation}
where \(\mathcal{E}\) is the conserved energy associated with the coordinate \(t\) (often interpreted as the total energy of the system in the context of particle motion), and \(L\) is the conserved angular momentum associated with the coordinate \(\phi\). These quantities are constant along the geodesic, and their conservation reflects the symmetries of the spacetime. The Lagrangian can now be rewritten in terms of \(\mathcal{E}\) and \(L\) as follows:
\begin{equation}
\mathcal{L} = -\mathcal{E}^2 + \dot{u}^2 + \frac{L^2}{\ell^2e^{\frac{2u}{R}}}.
\end{equation}
This reformulation of the Lagrangian simplifies the analysis of the motion along the radial coordinate \(u\), as the dynamics of the other coordinates have been encapsulated in the constants of motion \(\mathcal{E}\) and \(L\). In the case of null geodesics, where \(\kappa = 0\), the motion of the light ray is governed by the following equation (see also \cite{7,8}):
\begin{equation}
\left(\frac{du}{d\lambda}\right)^2 = \mathcal{E}^2 - \frac{L^2}{\ell^2e^{\frac{2u}{R}}}.
\end{equation}
This equation describes the radial motion of a photon under the influence of an effective potential, which depends on the energy \(\mathcal{E}\) and the angular momentum \(L\). The term \(\frac{L^2}{\ell^2e^{\frac{2u}{R}}}\) represents the effective potential, which dictates the behavior of the radial motion. This potential is given by (see also Figure \ref{fig:eff-pot-1}):
\begin{equation}
V_{\text{eff}}(u) = \frac{L^2}{\ell^2e^{\frac{2u}{R}}}.\label{eff-pot}
\end{equation}

This effective potential describes the interaction between the light ray and the spacetime geometry. The potential is always repulsive, meaning that it resists the inward motion of the photon. This repulsive force arises due to the curvature of the spacetime, and it pushes the particle away from the center of the geometry. At \(u=0\), the potential is constant: \(V_{\text{eff}}(u) = \frac{L^2}{\ell^2}\). This constant value represents a threshold that determines whether the photon can move inward. If the energy \(\mathcal{E}\) of the photon is lower than this threshold, the repulsive force of the potential will prevent it from moving inward. In this case, the photon will be "bounced back" by the potential. However, if the energy exceeds this threshold, the repulsive potential is overcome, and the photon can continue moving outward, freely passing through the region near the wormhole throat. As the radial coordinate \(u\) increases, the effective potential decays exponentially, and the repulsive force weakens. The decay is controlled by the parameter \(R\), which characterizes the size of the wormhole and the rate at which the potential diminishes with increasing \(u\). At large values of \(u\), the potential approaches zero, meaning that the spacetime no longer significantly influences the motion of the photon. At sufficiently large \(u\), the ray behaves like a free particle in flat spacetime, unaffected by the potential. The wormhole parameter \(\ell\) controls the strength of the effective potential near the throat. Smaller values of \(\ell\) correspond to a stronger repulsive potential, which makes it more difficult for the photon to move inward, while larger values of \(\ell\) lead to a weaker repulsive force, allowing for more freedom of motion. The parameter \(R\) governs the rate at which the potential decays with distance.

\vspace{0.15cm}
\setlength{\parindent}{0pt}

\begin{figure}[ht]
\centering
\includegraphics[scale=0.50]{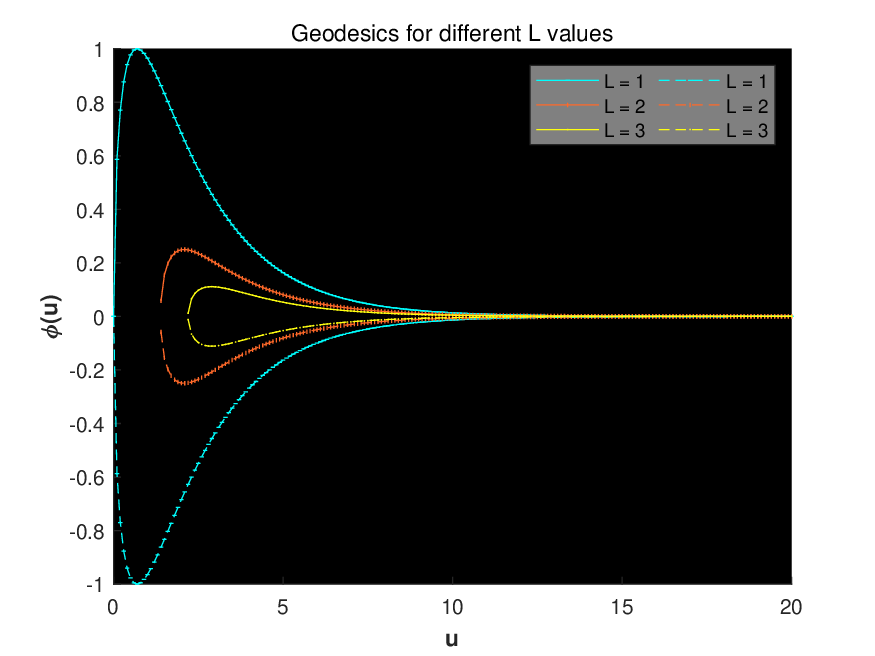}
\caption{\footnotesize Geodesics \(\phi(u)\) for varying values of \(L = 1, 2, 3\) over the range \(u \in [0, 20]\), with the initial condition \(\phi(u_i) = 0\). The solid and dashed lines represent the positive and negative branches of the geodesic, respectively. The calculations use the following parameters: \(R = 2\), \(\ell = 1\), \(\mathcal{E} = 1\), and \(\kappa = 0\).}
\label{fig:geodesics}
\end{figure}

Now, considering arbitrary geodesics, we obtain the following expression for the angular velocity:
\begin{equation}
\dot{\phi} = \frac{1}{\ell e^{\frac{u}{R}}\sqrt{\frac{(\mathcal{E}^2-\kappa)}{L^2} \, \ell^2e^{\frac{2u}{R}} - 1}} \, \dot{u}.
\end{equation}
This equation relates the angular velocity of the particle or photon to the radial velocity \(\dot{u}\) and the other conserved quantities, \(\mathcal{E}\) and \(L\). By integrating this expression, we can find the angular trajectory \(\phi(u)\), which gives the complete description of the geodesic motion in the spacetime. Accordingly, one finds:
\begin{equation}
\phi(u) = \phi(u_i) \pm \int_{u_i}^{u} \frac{du}{\ell e^{\frac{u}{R}}\sqrt{\frac{(\mathcal{E}^2-\kappa)}{L^2} \, \ell^2e^{\frac{2u}{R}} - 1}}.
\end{equation}
This integral describes the evolution of the angular coordinate \(\phi\) along the radial coordinate \(u\). The integration limits \(u_i\) and \(u\) correspond to the initial and final positions of the particle along the radial direction. Finally, by performing the integral explicitly, we obtain the following exact result for the angular coordinate:
\begin{equation}
\phi(u)= \phi(u_i) \pm \frac{R}{L\,\ell\, e^{\frac{u}{R}}}\sqrt{(\mathcal{E}^2-\kappa)\,\ell^2\, e^{\frac{2u}{R}}-L^2}.
\end{equation}
This expression provides a complete description of the angular evolution of the geodesic as a function of the radial coordinate \(u\), and it encapsulates the behavior of arbitrary geodesics in this particular spacetime geometry. The null geodesics for different values of $L$ can be seen in the Figure \ref{fig:geodesics}.

\section{\mdseries{Wave optics}}\label{sec:4}

\begin{figure}[ht]
\centering
\includegraphics[scale=0.50]{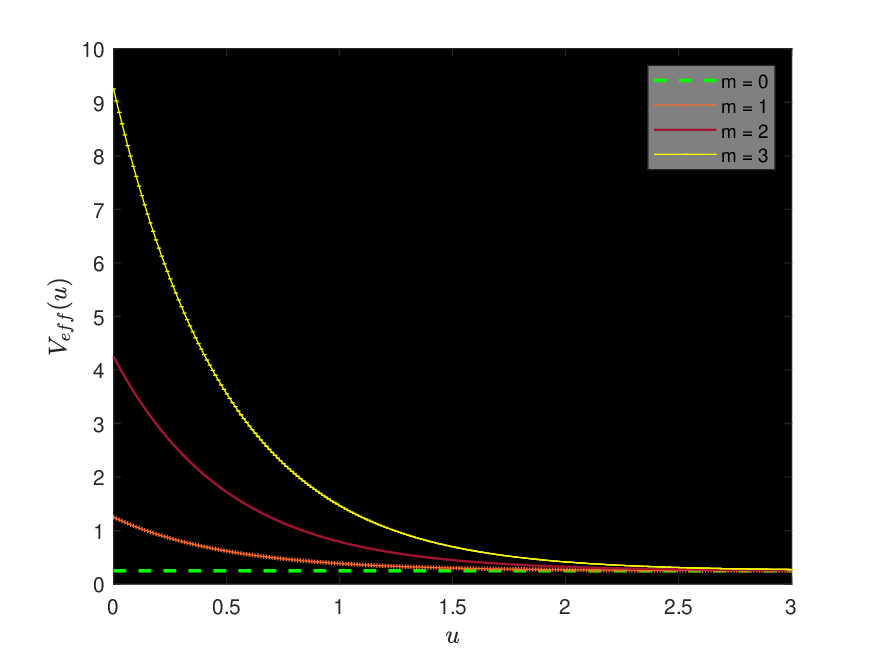}
\caption{\footnotesize Plot of the effective potential \( V_{\text{eff}}(u) \) as a function of \( u \) for various values of the angular momentum number \( m \), with fixed parameters \( R = 1 \) and \( \ell = 1 \). The potential decays exponentially with \( u \) and shows dependence on \( m \). The curves represent \( m = 0, 1, 2, 3 \), illustrating the effect of magnetic quantum number on the potential's shape and magnitude.}
\label{fig:eff-pot-2}
\end{figure}

In this section, we investigate the dynamics of wave propagation within the effective geometry of the Beltrami wormhole, focusing on the corresponding electromagnetic wave modes. In free space, Maxwell's equations \((\nabla \cdot \mathbf{E} = 0, \nabla \cdot \mathbf{B} = 0, \nabla \times \mathbf{E} = -\partial \mathbf{B}/\partial t, \nabla \times \mathbf{B} = \mu_0 \epsilon_0 \partial \mathbf{E}/\partial t)\) describe the electric \((\mathbf{E})\) and magnetic \((\mathbf{B})\) fields. Considering a monochromatic wave with time dependence \(e^{-i\omega t}\), starting from Maxwell's curl equations \((\nabla \times \mathbf{E} = -\mu_0 \partial \mathbf{B}/\partial t, \nabla \times \mathbf{B} = \epsilon_0 \partial \mathbf{E}/\partial t)\) and taking the curl of Faraday's law, we use the vector identity \(\nabla \times (\nabla \times \mathbf{E}) = \nabla(\nabla \cdot \mathbf{E}) - \nabla^2 \mathbf{E}\) along with \(\nabla \cdot \mathbf{E} = 0\) to obtain \(-\nabla^2 \mathbf{E} = -\mu_0 \epsilon_0 \partial^2 \mathbf{E}/\partial t^2\). Employing \(c^2 = 1/(\mu_0 \epsilon_0)\) leads to the Helmholtz equation \(\nabla^2 \mathbf{E} + (\omega^2/c^2)\mathbf{E} = 0\), or \(\nabla^2 \psi + k^2 \psi = 0\) with wave number \(k = \omega/c\). This equation captures the spatial behavior of Maxwell's equations under time-harmonic conditions, describing electromagnetic wave propagation in a source-free region, with its solutions representing the allowed electromagnetic wave modes in the frequency domain. The Helmholtz equation, which governs wave propagation in a curved spacetime or geometry, is essential for our analysis. To begin with, we write the Helmholtz equation within this specific geometry \cite{8}:
\begin{equation}
    (\Delta_{g} + k^2)\Psi = 0,
\end{equation}
where \( \Delta_{g} \) represents the Laplace-Beltrami operator. The term \( k^2 \) corresponds to the propagation constant associated with the waves in the given geometry, while \( \Delta_{g}\) encapsulates the effects of curvature on wave propagation. Specifically, for a generic metric \( ds^2 = g_{ij} dx^i dx^j \), the Laplace-Beltrami operator takes the following form \cite{8}:
\begin{equation}
    \Delta_{g} \Psi = \frac{1}{\sqrt{g}} \partial_i \left( \sqrt{g} g^{ij} \partial_j \Psi \right),
\end{equation}
where \( g = |\text{det}(g_{ij})| \) is the determinant of the metric tensor \( g_{ij} \), and the indices \( i, j \) correspond to the coordinates in the geometry, which in this case are \( u \) and \( \phi \). To proceed, we now express the Helmholtz equation in a form suitable for the specific geometry of the Beltrami wormhole. According to Eq. \eqref{metric}, we obtain the following modified form of the Helmholtz equation for the wave function \( \Psi(u, \phi) = \psi(u)\,e^{i m \phi} \), with \( m=0,\pm 1, \pm 2... \) being the magnetic quantum number, given the periodic boundary condition on \( \phi \):
\begin{equation}
\left[ \partial^2_{u} + \frac{1}{R}\,\partial_{u} + k^2 - \frac{m^2}{\ell^2 e^{\frac{2u}{R}}} \right] \psi(u) = 0. \label{WE}
\end{equation}
Now, let us try to determine the effective potential acting on waves within the curved background. To do this, we remove the first-order derivative term, $\frac{1}{R}\,\partial_{u} \psi(u)$. This can be achieved via a transformation of the radial wave function, \( \psi(u) \), as follows: \(\psi(u) = e^{-\frac{u}{2R}} \varphi(u)\). After applying this transformation, the wave equation simplifies to a one-dimensional Schrödinger-like equation:
\begin{equation}
\ddot{\varphi}(u) + \left[k^2 - V_{\text{eff}}(u)\right] \varphi(u) = 0, \label{Sch-type}
\end{equation}
where \( V_{\text{eff}}(u) \) represents the effective potential governing wave propagation in the wormhole geometry and is given by (see also Figure \ref{fig:eff-pot-2}):
\begin{equation}
V_{\text{eff}}(u) = \frac{m^2}{\ell^2 e^{\frac{2u}{R}}}+\frac{1}{4R^2} . \label{eff-pot2}
\end{equation}

\begin{figure*}[h!]
\centering
\includegraphics[scale=0.60]{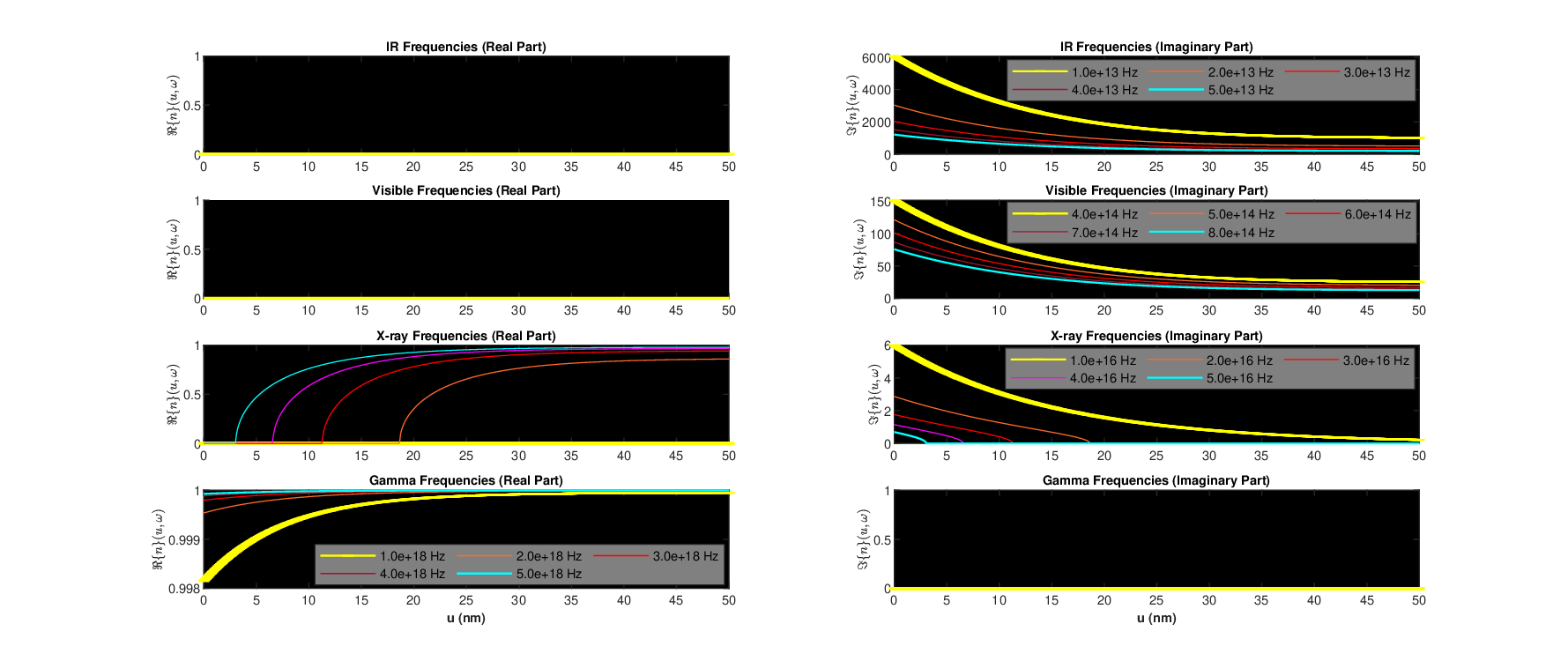}
\caption{ \footnotesize Real and imaginary parts of the refractive index \( n(u, \omega) \) as functions of the parameter \( u \) for different electromagnetic frequency regions: infrared (IR), visible light, X-ray, and gamma rays. The calculations are performed using the parameters \( R = 15 \) nm, \( \ell = 5 \) nm, \( m = 1 \), and the speed of light \( c = 3 \times 10^{17} \) nm/s. Each row corresponds to a specific frequency region, with the left column displaying the real part \( \Re\{n\} \) and the right column showing the imaginary part \( \Im\{n\} \). The refractive index is computed for five representative frequencies within each region. These plots illustrate variations in optical properties across different wavelengths, highlighting how light interacts with the medium in each spectral range.}
\label{fig:refractive-index}
\end{figure*}

The effective potential (\ref{eff-pot2}) is crucial for understanding how waves are affected by the geometry of the Beltrami wormhole. It depends on the radial coordinate \( u \), the magnetic quantum number \( m \), and the geometry parameters \( R \) and \( \ell \). Obviously, the Beltrami wormhole spacetime manifestly introduces a repulsive core (i.e., effective potential) even for $m=0$. The repulsion strength of which decreases exponentially with increasing $u$ and stabilizes at the constant value $\frac{1}{4R^2}\neq0$ as $u\rightarrow\infty$. In the vicinity of the wormhole throat ($u\rightarrow0$), on the other hand, the strength of the repulsive core is dominated by \(\frac{m^2}{\ell^2}+\frac{1}{4R^2}\). Moreover, the effect of the scaling factor \( R \) in \(\frac{1}{4R^2}\) is insignificant and cannot compete with the rapidity/speed of decay in the exponential term \(\sim \exp(-2u/R)\) as $u\rightarrow\infty$. The very nature of the repulsive gravitational force, introduced by the curved spacetime, suggests that the propagation of the waves is faster near the wormhole throat, but becomes slower as the wave propagates far from the wormhole throat. This is attributed to the strength of the repulsive core, which is maximum at \(u\sim 0\) and minimum at \(u\sim \infty\). These results suggest a space-dependent wave number and may allow us to determine a space- and frequency-dependent refractive index, \(n(u, \omega)\). Let us now derive the expression for \(n(u, \omega)\). In the presence of a spatially varying medium, the Helmholtz equation can be expressed as \(
\nabla^2 \varphi + k_{\text{eff}}^2(u) \varphi = 0\), where \(k_{\text{eff}}(u)\) is the effective wave number that accounts for the medium's influence. The refractive index of the medium is related to the effective wave number by \(k_{\text{eff}}(u) = \frac{\omega}{c} n(u)\). Substituting this relation into the Helmholtz equation yields \(\nabla^2 \varphi + \left(\frac{\omega^2}{c^2} n^2(u)\right) \varphi = 0\). This is the general form of the wave equation in a medium with a spatially varying refractive index. By recognizing that the term \((k^2 - V_{\text{eff}}(u))\) modifies the system's wave number and comparing it to the wave equation in a refractive index medium, we obtain \(k_{\text{eff}}^2(u) = k^2 - V_{\text{eff}}(u)\), which leads to \(n^2(u) = 1 - \frac{c^2 V_{\text{eff}}(u)}{\omega^2}\). Accordingly, the refractive index is given by (see also Figures \ref{fig:refractive-index} and \ref{fig:ref-index-2}):
\begin{equation}
n(u,\omega) = \sqrt{1 - \frac{c^2 m^2}{\omega^2 \ell^2 e^{\frac{2u}{R}}} - \frac{c^2}{4\omega^2 R^2}}.\label{Ref-indeks}
\end{equation}
This expression clearly shows that the refractive index depends on the parameters of the effective potential as well as the wave frequency. The refractive index can be less than 1 or even imaginary, depending on these parameters. A refractive index \(n < 1\) indicates that the phase velocity of light or waves in this medium exceeds $c$. This characteristic is a hallmark of metamaterials artificially engineered materials exhibiting unconventional optical properties, including negative refraction, where light bends opposite to the direction observed in conventional materials. Metamaterials with \(n< 1\) can enable phenomena like superlensing, which surpasses the diffraction limit to produce high-resolution images. Notably, a refractive index smaller than 1, leading to superluminal phase velocities, does not violate causality or special relativity since the group velocity, responsible for information transfer, remains constrained by $c$.

\section{\mdseries{Summary and discussions}}\label{sec:5}

In this paper, we investigate the behavior of ray trajectories and wave dynamics on a Beltrami surface, also known as a Beltrami wormhole or optical condensed matter wormhole. Using the Lagrangian formalism, we derive exact angular trajectories as functions of the radial optical distance, providing valuable insights into the geodesic motion of particles and light within this curved spacetime structure. The Lagrangian approach offers a fundamental description of particle trajectories along geodesics, with the affine parameter \(\lambda\). The nature of the geodesics is determined by the constant \(\kappa\), which distinguishes light-like paths (\(\kappa = 0\)) suitable for photons from time-like paths (\(\kappa = -1\)) associated with massive particles. By recognizing \(t\) and \(\phi\) as cyclic coordinates, we obtain conserved quantities: energy \(\mathcal{E} = \dot{t}\) and angular momentum \(L = \ell^2 e^{\frac{2u}{R}} \dot{\phi}\) (with \(\ell \neq 0\)), which govern the dynamics of the angular trajectory. This shows the direct connection between symmetry and conserved motion in the context of geodesic flow.

\begin{figure}[h!]
\centering
\includegraphics[scale=0.60]{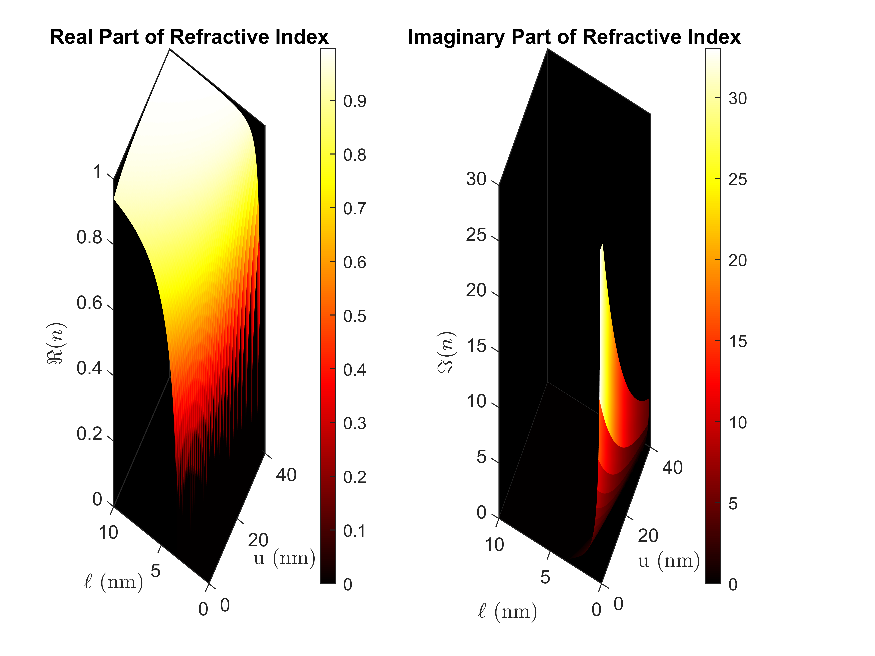}
\caption{ \footnotesize 3D surface plots of the real and imaginary parts of the refractive index \( n(u, \omega) \) as functions of the parameter \( u \) and wormhole throat \( \ell \) for a fixed X-ray frequency (\(\omega = 3 \times 10^{17}\) Hz). The calculations are performed with the speed of light set to \( c = 3 \times 10^{17} \) nm/s, a fixed radius \( R = 20 \) nm, and magnetic quantum number \( m = 1 \). The left subplot illustrates the real part \( \Re\{n\} \), while the right subplot shows the imaginary part \( \Im\{n\} \). The refractive index is computed for varying \( u \) in the range of 0 to 40 nm and \( \ell \) from 1 to 10 nm. These plots demonstrate the dependence of optical properties on geometric parameters in the X-ray regime.}
\label{fig:ref-index-2}
\end{figure}

\vspace{0.15cm}
\setlength{\parindent}{0pt}

The effective potential \(V_{\text{eff}}(u) = \frac{L^2}{\ell^2 e^{\frac{2u}{R}}}\), arising from the curvature of spacetime, introduces a repulsive force that influences radial motion. This repulsion becomes particularly significant near the wormhole throat, preventing inward motion unless the photon’s energy exceeds a critical threshold as \(u \to 0\). The potential decays exponentially with increasing radial optical distance \(u\), reflecting the diminishing influence of spacetime curvature, governed by the parameter \(R\). The effect of increasing angular momentum \(L\) is to strengthen the repulsive potential, which alters the trajectories. Larger values of \(R\) result in a slower decay of the potential, thus extending the wormhole’s influence.

\vspace{0.15cm}
\setlength{\parindent}{0pt}

Wave propagation within the Beltrami wormhole geometry can be analyzed by modifying the Helmholtz equation for curved spacetime. The resulting Schrödinger-like equation features an effective potential with a centrifugal barrier \(\frac{m^2}{\ell^2 e^{\frac{2u}{R}}}\) and a constant term \(\frac{1}{4R^2}\). The centrifugal barrier, which plays a crucial role near the wormhole throat, decays exponentially with \(u\), while the constant term ensures a baseline repulsive force. This baseline force can prevent free wave propagation under certain conditions determined by the wave's energy. For \(m = 0\), the potential is dominated by the constant term, maintaining a persistent repulsive effect. Larger values of \(R\) slow the decay of the centrifugal barrier, extending its influence over greater distances, which effectively inhibits free wave propagation in this wormhole-like spacetime.

\vspace{0.15cm}
\setlength{\parindent}{0pt}

The refractive index on a Beltrami surface is given in \eqref{Ref-indeks}, and it exhibits distinct characteristics that depend on the wave frequency \(\omega\). For \(0 < n(u) < 1\), the condition \(\omega > c \sqrt{\frac{m^2}{\ell^2 e^{\frac{2u}{R}}} + \frac{1}{4R^2}}\) must hold. The relationship between the wormhole throat radius \(\ell\) and the curvature scale \(R\) strongly influences the refractive index, leading to surface-bound waves with phase velocities greater than \(c\), as found in geometric waveguides, while still maintaining causality. Near the wormhole throat, the refractive index approaches unity, indicating that the magnetic quantum number \(m\), geometric curvature, and wave frequency collectively govern the optical properties of the surface. In optical and wave physics, the refractive index \( n \) dictates the fundamental interaction between electromagnetic waves and a medium, governing both the phase velocity and attenuation properties. A spatially varying refractive index modifies local dispersion relations, leading to complex wave dynamics, including nontrivial refraction, gradient-induced localization, and engineered anisotropy \cite{30}. When \( 0 < n < 1 \), the phase velocity of light exceeds \( c \), a phenomenon characteristic of low-density plasmas, X-ray propagation in weakly polarizable media, and certain metamaterial structures. In plasma physics, this behavior follows from the refractive index expression \( n = \sqrt{1 - \omega_p^2 / \omega^2} \), where \( \omega_p \) is the plasma frequency. While the phase velocity in this regime is superluminal, causality remains preserved since the group velocity, associated with energy transport, is always subluminal. The implications of \( 0 < n < 1 \) extend to transformation optics, where refractive index gradients enable precise wavefront manipulation \cite{30}, and to relativistic quantum field theory, where analogous dispersion relations arise in effective metric descriptions of wave propagation in curved spacetime.

For frequencies satisfying \(\omega < c \sqrt{\frac{1}{4R^2} + \frac{m^2}{\ell^2 e^{\frac{2u}{R}}}}\), the expression under the square root becomes negative, leading to an imaginary refractive index. A purely imaginary refractive index, defined as \( n = i\tilde{\kappa} \) with \( \tilde{\kappa} > 0 \), corresponds to a regime of extreme wave attenuation rather than propagation \cite{31,32}. In this regime, the complex wave vector induces an exponential decay of the field, characterized by a penetration depth \( \delta \sim 1/(\tilde{\kappa} k) \), where \( k \) is the free-space wave number. Such behavior is observed in highly lossy materials, including metals at optical frequencies, where strong electron scattering results in significant absorption, as well as in total internal reflection beyond the critical angle, where evanescent waves decay away from the interface \cite{31,32}. The frequency range in which the refractive index becomes imaginary gives rise to a bandgap-like effect, prohibiting the propagation of certain frequencies and leading to the formation of surface-bound or localized modes. The onset of this attenuation is determined by the magnetic quantum number \( m \), the wormhole throat radius \( \ell \), and the curvature scale \( R \), exhibiting similarities to plasmonic systems and optical materials with strong absorption. For S-waves (\( m = 0, \ell = 0 \)), the refractive index remains real and positive, allowing wave propagation when the frequency exceeds the threshold \(\omega > \frac{c}{2R}\). Below this threshold (see also \cite{33}), the refractive index becomes imaginary, causing the waves to transition into evanescent modes and decay exponentially.

Figures \ref{fig:refractive-index} and \ref{fig:ref-index-2} illustrate how the optical properties vary as a function of frequency and the geometric parameters \( R \) and \( \ell \). These results may have significant implications for understanding light and wave behavior in graphene-like condensed matter systems, where spacetime curvature plays a crucial role in light propagation. Furthermore, these findings provide a theoretical framework for designing novel optical systems, such as waveguides and lenses, that exploit spacetime curvature for applications in quantum optics and photonics. In principle, by controlling wave dynamics in curved geometries, we can advance technologies that utilize the unique effects of spacetime curvature, particularly in advanced nanomaterials.

\section*{\small{CRediT authorship contribution statement}}

\section*{\small{Data availability}}

The authors confirm that the data supporting the findings of this study are available within the article.

\section*{\small{Conflicts of interest statement}}

The authors have disclosed no conflicts of interest.

\section*{\small{Funding}}

This research has not received any funding.

\end{document}